\documentclass[12pt]{article}
\title{
Thermal Equilibrium of a Macroscopic Quantum System in a Pure State}
\author{
Sheldon Goldstein\footnote{Department of Mathematics,
     Rutgers University, Hill Center,
     110 Frelinghuysen Road, Piscataway, NJ 08854-8019, USA.
     E-mail: oldstein@math.rutgers.edu},
David A. Huse\footnote{Department of Physics, Princeton University,
    Jadwin Hall, Washington Road, Princeton, NJ  08544-0708, USA.
    E-mail: huse@princeton.edu},\\
Joel L. Lebowitz\footnote{Departments of Mathematics and Physics,
     Rutgers University, Hill Center,
     110 Frelinghuysen Road, Piscataway, NJ 08854-8019, USA.
     E-mail: lebowitz@math.rutgers.edu}, and
Roderich Tumulka\footnote{Department of Mathematics,
     Rutgers University, Hill Center,
     110 Frelinghuysen Road, Piscataway, NJ 08854-8019, USA.
     E-mail: tumulka@math.rutgers.edu}
}
\date{June 23, 2015}

\usepackage{amsthm,url,amssymb,amsmath,amsfonts,mathrsfs}

\newcommand{\ket}[1]{\vert#1\rangle}
\newcommand{\bra}[1]{\langle#1\vert}
\newcommand{\pr}[1]{\ket{#1}\bra{#1}}
\DeclareMathOperator{\tr}{tr}
\newcommand{\MATE}{\mathrm{MATE}}
\newcommand{\TMATE}{\mathrm{TMATE}}
\newcommand{\MITE}{\mathrm{MITE}}

\newcommand{\RRR}{\mathbb{R}}

\newcommand{\SSS}{\mathbb{S}}

\newcommand{\scp}[2]{\langle #1| #2 \rangle}
\newcommand{\Hilbert}{\mathscr{H}}

\newcommand{\be}{\begin{equation}}
\newcommand{\ee}{\end{equation}}
\newcommand{\mc}{\mathrm{mc}}
\newcommand{\eq}{\mathrm{eq}}
\newcommand{\sA}{\mathscr{A}}
\newcommand{\length}{L}

\begin{document}
\maketitle
\begin{abstract}
We consider the notion of thermal equilibrium for an \emph{individual}
closed macroscopic quantum system in a pure state, i.e., described by a wave
function. The macroscopic properties in thermal equilibrium of such a system, determined by its
wave function, must be the same as those obtained from thermodynamics,
e.g., spatial uniformity of temperature and chemical potential. When this
is true we say that the system is in \emph{macroscopic thermal equilibrium}
(MATE). Such a system may however not be in \emph{microscopic thermal
equilibrium} (MITE). The latter requires that the reduced density matrices
of small subsystems be close to those obtained from the microcanonical,
equivalently the canonical, ensemble for the \emph{whole} system. The
distinction between MITE and MATE is particularly relevant for systems
with many-body localization (MBL) for which the energy eigenfuctions fail
to be in MITE while necessarily most of them, but not all, are in MATE.
We note however that for generic macroscopic systems, including those
with MBL, most wave functions in an energy shell are in both MATE and
MITE. 
For a classical macroscopic system, MATE holds for most phase points on the energy surface, but MITE fails to hold for any phase point.

\medskip

\end{abstract}

\section{Introduction}
\label{sec:intro}

Thermal behavior of closed macroscopic systems in pure states has been widely studied in recent years; see e.g., \cite{GMM04,PSW06,GLTZ06,Sug07,Rei07,Rei08,RDO08,LPSW08,GLMTZ09b,Rei10,RS12,Rei15,GE15}, after some pioneering work even earlier \cite{Schr27,vN29,Schr52,Deu91,Sre94,T98}.
In particular, the importance of the \emph{eigenstate thermalization
hypothesis} (ETH) \cite{Deu91,Sre94} has become widely appreciated, see e.g., \cite{RDO08,Rei08,LPSW08,GLMTZ09b,GLTZ10,NH15,Rei15,GE15}. It 
asserts, in one version, that all energy
eigenstates (in a suitable energy shell) are thermal, i.e., assigning probability distributions to observables that are characteristic of thermal equilibrium. 
The ETH holds for many (but not all) macroscopic quantum systems. 
When it holds then the system, starting out of equilibrium,
will thermalize (at least in the time average; see Section~\ref{sec:discuss} below). Thus if a system does not thermalize it must have
energy eigenfunctions that fail to be thermal.

An important case of this is that of many-body localization (MBL) \cite{And58,BAA06b,OH07}, for
which the Hamiltonian has (at least some) eigenfunctions that are in some way localized, so that, for any wave function, the component from these eigenfunctions will not spread but stay localized forever. For systems with MBL it has been argued that most (if not
all) energy eigenfunctions (in suitable energy intervals) fail to be
thermal, and there are in fact models for which this can be analytically \cite{Imbrie}, numerically \cite{PH10}, or perturbatively \cite{BAA06a,BAA06b,ros}
seen to be the case. At the same time it has been argued \cite{GLMTZ09b} that most
energy eigenstates must rather generally be thermal, in particular even
for systems with MBL.

This situation is thus rather puzzling: How can the fact that most
energy eigenfunctions for (some) MBL systems
are not  thermal be reconciled with the results in \cite{GLMTZ09b} showing
that in wide generality most of them must be.
The answer to this question, we point out, lies in the fact that there
are basically two notions of thermal equilibrium for pure
states: a macroscopic notion of thermal equilibrium that  we call MATE,
and a more refined microscopic one that  we
call MITE. While most, but definitely not all, energy eigenstates
of a system with MBL
are in MATE (see Sec.~\ref{sec:discuss}), none, or nearly none, are in MITE \cite{BAA06a,BAA06b,PH10,Imbrie}.
Nonetheless, most pure states in the energy shell are in both MATE and
MITE, even for systems with MBL (see Sec.~\ref{sec:MATE} and \ref{sec:MITE}).
It is with these two notions of thermal equilibrium and their consequences
that we are concerned here.

To be more precise, consider a finite, macroscopic, closed quantum system with Hilbert space $\Hilbert$. Let $\Hilbert_{\mc}$ be a micro-canonical energy shell, i.e., the
subspace of $\Hilbert$ spanned by the energy eigenstates with eigenvalue in an energy interval that is small on the macroscopic scale but contains many eigenvalues. The micro-canonical density matrix $\hat\rho^{\mc}$ is defined by $\hat\rho^{\mc} = (\dim\Hilbert_{\mc})^{-1} \hat{P}_{\mc}$ with $\hat{P}_{\mc}$ the projection to $\Hilbert_{\mc}$. As usual, pure states in $\Hilbert_{\mc}$ are superpositions of energy eigenstates in $\Hilbert_{\mc}$.  Both MITE and MATE can be expressed as subsets of the unit sphere in $\Hilbert_{\mc}$,
\be
\SSS(\Hilbert_{\mc})=\{\psi\in\Hilbert_{\mc}:\|\psi\|=1\}\,.
\ee

We will often simply say ``micro'' for ``microscopic'' and ``macro'' for ``macroscopic''; ``most'' means ``all but a few'' or ``all except a set of small measure'' (i.e., ``an overwhelming majority of''); measures are taken to be normalized; the small measure, in fact, tends to 0 in the thermodynamic limit. We define MATE and MITE in Sections~\ref{sec:MATE} and \ref{sec:MITE}, respectively, and discuss the properties and differences of the two notions in Section~\ref{sec:discuss}. We also note there that these notions can also be applied to mixed states. We provide a deeper and more detailed discussion elsewhere \cite{GHLT15b}.

\section{MATE}
\label{sec:MATE}

The definition of MATE is based on macro observables $\hat{M}_1,\ldots,\hat{M}_K$. To be specific, macro observables can reasonably be based on a 
partition of the system's available volume $\Lambda\subset \RRR^3$ into cells $\Lambda_i$ that are small on the macro scale but 
still large enough to each contain a large number of degrees of freedom. Examples of natural choices of  $\hat M$'s are, for each cell, the number of 
particles of each type, the total energy, the total momentum, and/or the total magnetization. These are the variables usually considered in the thermodynamic or hydrodynamic description of a fluid or magnet.

Following von Neumann \cite{vN29}, we take the $\hat M_j$ to commute with each other and to be such that the gaps between the eigenvalues are of the order of the macroscopic resolution (so that the eigenvalues are highly degenerate). This can be achieved by suitably ``rounding off'' and coarse-graining the operators representing the macro observables \cite{vN29,GLTZ10,Y}. Taking $\Hilbert_{\mc}$ to be an eigenspace of a ``macro energy'' operator, and thus to commute with the other macro observables, all $\hat{M}_j$ can be regarded as operators on $\Hilbert_{\mc}$. Their joint spectral decomposition defines an orthogonal decomposition
\be
\Hilbert_{\mc} = \bigoplus_\nu \Hilbert_\nu\,,
\ee
and the subspaces $\Hilbert_{\nu}$ (``macro spaces''), the joint eigenspaces of the macro observables, correspond to the different macro states and have very high dimension. The decomposition is in some ways analogous to a partition, in classical mechanics, of an energy shell $\Gamma_{\mc}$ 
in phase space into disjoint subsets $\Gamma_\nu$ corresponding to different macro states \cite{B96,Gol99,GL,L07}. It is generally the 
case \cite{B96,Lan,GL}, both in classical and quantum mechanics, that one of the macro states,  corresponding to thermal equilibrium, is dominant, i.e., 
that one of the $\Gamma_\nu$'s, denoted $\Gamma_{\eq}$, has 
most of the phase space volume of $\Gamma_{\mc}$, and that one of the $\Hilbert_\nu$, denoted $\Hilbert_{\eq}$, has most 
of the dimensions of $\Hilbert_{\mc}$, i.e.,
\be\label{dimeqdimmc}
\frac{\dim\Hilbert_{\eq}}{\dim \Hilbert_{\mc}} = 1-\varepsilon
\ee
with $\varepsilon\ll 1$.\footnote{An exception to the existence of a dominant macro space is provided by first-order phase transitions, such as in the ferromagnetic Ising model in a vanishing external magnetic field, where $\Hilbert_\nu$ has the appropriate majority of spins up and $\Hilbert_{\nu'}$ has the appropriate majority of spins down, each having nearly 50\%\ of the dimension of $\Hilbert_{\mc}$ for a suitable energy interval.}  Realistic values of $\varepsilon$, say for a liter of air under atmospheric pressure, are smaller than $10^{-10^5}$ \cite{GHLT15b}; more generally, $\varepsilon$ is exponentially small in the number of degrees of freedom per cell. We assume here that our system and our choice of macro observables are such that \eqref{dimeqdimmc} holds for suitably small $\varepsilon$.

The system is said to be in \emph{MATE} whenever its wave function $\psi$ lies within a $\delta$-neighborhood of $\Hilbert_{\eq}$ with suitably small $\delta>0$, i.e., in the set
\be\label{MATEdef}
\MATE = \Bigl\{ \psi\in\SSS(\Hilbert_{\mc}): \scp{\psi}{\hat P_{\eq}|\psi} >1- \delta \Bigr\}
\ee
with $\hat P_{\eq}$ the projection to $\Hilbert_{\eq}$. 
Thus for a state $\psi$ that is in MATE, the probability is close to one that all macro observables take on their thermal equilibrium values. 
A concept of thermal equilibrium along these lines was used before in, e.g., \cite{Gri,RDO08,GLMTZ09b,GHT13b,GHT14a,GHT14b}.
It is known \cite{GLMTZ09b} that, if $\varepsilon\ll \delta$, then $\MATE$ has most 
of the surface area of $\SSS(\Hilbert_{\mc})$, so most pure states are in MATE. This follows from the fact that the average of $\scp{\psi}{\hat P_{\eq}|\psi}$ over $\SSS(\Hilbert_{\mc})$ is equal to
\eqref{dimeqdimmc}; since $\scp{\psi}{\hat P_{\eq}|\psi}$ cannot exceed 1, it must be close to 1 for most $\psi$ (see Section~\ref{sec:discuss} for more detail). 
It can similarly be shown, see \eqref{mosteigenstates} below, that most energy eigenstates are in MATE.
To be sure, there are states in the energy shell which are not in MATE; for example, one could take a tensor product of states of two regions having (what look macroscopically like) different temperatures.

An alternative definition due to Tasaki \cite{Tas15} (in the same direction as \cite{DRMN06,Sug07}), not strictly but approximately equivalent and denoted TMATE here, avoids the step of rounding off to make the macro observables commute, which may pose substantial difficulty to carry out in practice. Instead, take $\hat{M}_1,\ldots,\hat{M}_K$ to be the macro observables before rounding off and coarse graining (mathematically, any self-adjoint operators), let $V_j=\tr(\hat{\rho}^{\mc} \, \hat{M}_j)$ be the thermal equilibrium value of $\hat{M}_j$, and let $\Delta M_j$ be the macro resolution of the observable represented by $\hat{M}_j$. Using $1_A$ to denote the characteristic function of the set $A$, we define
\be
\hat{P}_j = 1_{[V_j-\Delta M_j,V_j+\Delta M_j]}(\hat M_j)
\ee
to be the projection associated with the eigenvalues of $\hat{M_j}$ that lie within the macro resolution of the thermal equilibrium value. Then let
\be\label{TMATEdef}
\TMATE = \bigcap_{j=1}^K
\Bigl\{ \psi\in\SSS(\Hilbert_{\mc}): \scp{\psi}{\hat{P}_j|\psi}>1-\delta \Bigr\}\,.
\ee
Note that $\scp{\psi}{\hat{P}_j|\psi}$ is the probability of finding, in a quantum measurement of $\hat M_j$ on a system in state $\psi$, a value that is $\Delta M_j$-close to $V_j$. If this probability is $>1-\delta$ for at least the fraction $1-\eta$ of $\SSS(\Hilbert_{\mc})$ for each $j$, then $\TMATE$ has at least size $1-K\eta$ (in terms of normalized surface area), which is close to 1 if $\eta \ll K^{-1}$.
We note further that MATE as in \eqref{MATEdef} can essentially also be written as the right-hand side of \eqref{TMATEdef} if the $\hat M_j$ are taken again as commuting and coarse-grained on the scale $\Delta M_j$, so that $\hat P_j$ becomes the projection to the eigenspace of $\hat M_j$ with the dominant (most degenerate) eigenvalue, and $\Hilbert_{\eq}$ is the intersection of these eigenspaces. Indeed, the right-hand side of \eqref{TMATEdef} is then contained in that of \eqref{MATEdef} with $\delta$ replaced by $K\delta$, and that of \eqref{MATEdef} is then contained in that of \eqref{TMATEdef}.

\section{MITE}
\label{sec:MITE}

While MATE implies thermal behavior only for macro observables, MITE involves also ``micro'' observables, more precisely, those observables concerning only a region smaller than a certain length scale $\ell$. The definition of MITE is inspired by \emph{canonical typicality}, the observation \cite{GMM04,PSW05,PSW06,GLTZ06} that for any not-too-large subsystem $S$ and most wave functions $\psi$ in the energy shell $\Hilbert_{\mc}$, the reduced density matrix of $S$ is close to the thermal equilibrium density matrix of $S$,
\be\label{rhopsirhomc}
\hat\rho^\psi_S \approx \hat\rho^{\mc}_S\,,
\ee
where
\be\label{rhoSdef}
\hat\rho^\psi_S= \tr_{S^c} \pr{\psi}
\ee
is the reduced density matrix of $S$ obtained by tracing out the complement $S^c$ of $S$, and
\be
\hat\rho^{\mc}_S= \tr_{S^c} \hat\rho^{\mc}\,.
\ee
If $S$ is small enough then
\be\label{rhomcrhoc}
\hat\rho^{\mc}_S \approx \hat\rho^{(\beta)}_S
\ee
for suitable $\beta>0$, where the right-hand side is the partial trace,
\be\label{rhoc}
\hat\rho_S^{(\beta)}= \tr_{S^c} \, \hat\rho^{(\beta)}\,,
\ee
of the canonical density matrix
\be
\hat\rho^{(\beta)} = \frac{1}{Z} e^{-\beta \hat H} \text{ with }Z = \tr e^{-\beta\hat H}\,.
\ee
As a consequence, for small $S$, it does not matter whether one starts from $\hat\rho^{\mc}$ or $\hat\rho^{(\beta)}$ (this fact is a version of equivalence of ensembles), and we also have
\be
\hat\rho^\psi_S \approx \hat\rho^{(\beta)}_S\,.
\ee
We will call \eqref{rhoc} the canonical or thermal density matrix for $S$.\footnote{The density matrix $Z^{-1}_S\exp(-\beta \hat{H}_S)$ with $\hat H_S$ the Hamiltonian of $S$ 
is sometimes called the canonical or thermal density matrix for $S$; it agrees with \eqref{rhoc} if the interaction between $S$ and its complement can be neglected. If the interaction cannot be neglected, then \eqref{rhoc} is the correct density matrix to use.}
We note that if $\hat\rho^\psi_S \approx \hat\rho^{\mc}_S$ for some subsystem $S$ then the same is true for every smaller subsystem $S'$ contained in $S$ (``subsubsystem property'' of \eqref{rhopsirhomc}),
just by taking another partial trace on both sides of the approximate equation $\hat\rho^\psi_S \approx \hat\rho^{\mc}_S$.

The system is said to be in $\MITE_\ell$ (MITE on the length scale $\ell$) whenever $\psi\in\SSS(\Hilbert_{\mc})$ satisfies \eqref{rhopsirhomc} for every subsystem $S$ corresponding to a spatial region of diameter $\mathrm{diam}(S)\leq \ell$, i.e.,
\be\label{MITEdef}
\MITE_\ell = \bigcap_{S\: : \: \mathrm{diam}(S)\leq \ell} \Bigl\{ \psi\in\SSS(\Hilbert_{\mc}): \hat\rho^{\psi}_S \approx \hat\rho^{\mc}_S\Bigr\}
\ee
with some precise definition of $\approx$ (such as the trace norm of the difference being smaller than a given value). The subsubsystem property implies that every $\psi$ in $\MITE_\ell$ lies also in $\MITE_{\ell'}$ for any smaller scale $0<\ell'<\ell$.

\emph{MITE} is then defined to mean $\MITE_{\ell_0}$ with $\ell_0$ the largest $\ell$ small enough to ensure that \eqref{rhomcrhoc} holds for every subsystem $S$ 
with $\mathrm{diam}(S)\leq \ell_0$. As a practical value,
for example, we may take
\be\label{Lvalue}
\ell_0 =10^{-3} \, \mathrm{diam}(\Lambda)\,,
\ee
where $\Lambda\subset\RRR^3$ is the volume of the whole system. Thus a state $\psi$ is in MITE if for every subsystem of diameter $\ell_0$ or smaller, the reduced density matrix is close to the thermal equilibrium reduced density matrix.

Most $\psi\in\SSS(\Hilbert_{\mc})$ lie in MITE.
Indeed, canonical typicality (in the sense of $\hat\rho^\psi_S\approx \hat\rho^{\mc}_S$ for most $\psi$) holds for subsystems of up to nearly 
half the size of $\Lambda$ (that is, half the degrees of freedom, or square root of the Hilbert space dimension, in practice usually half of the volume; 
see \cite{PSW05,PSW06,GHLT15b} for a discussion). We can choose a moderate number $r$ (e.g., $r=8$ for cube-shaped $\Lambda$) of overlapping regions $S_i\subset\Lambda$ 
(e.g., also cubes) of nearly half the volume so that most $\psi$ satisfy $\hat\rho^\psi_{S_i}\approx \hat\rho^{\mc}_{S_i}$ for all $1\leq i\leq r$ 
simultaneously, and so that every region $S$ with $\mathrm{diam}(S)\leq \tfrac{1}{4}\,\mathrm{diam}(\Lambda)$ is contained in one of the $S_i$. 
By the subsubsystem property, also $\hat\rho^\psi_S\approx \hat\rho^{\mc}_S$ for such regions $S$, so most $\psi$ lie in $\MITE_{\mathrm{diam}(\Lambda)/4}$ and 
a fortiori in $\MITE=\MITE_{\ell_0}$ with $\ell_0$ as in \eqref{Lvalue}.

A concept along the lines of MITE was used before in, e.g., \cite{Rei08,LPSW08,RS12,NH15}.

\section{Discussion}
\label{sec:discuss}

We organize our discussion under a number of subheadings.

{\it MITE implies MATE.}---To discuss this point, it is helpful to first introduce a common framework for formulating MITE and MATE. For any observable $\hat A$, let $\mu^\psi_{\hat A}$ denote the probability distribution defined by $\psi\in\SSS(\Hilbert_{\mc})$ over the spectrum of $\hat A$,
\be
\mu^\psi_{\hat A} (B) = \scp{\psi}{1_B(\hat A)|\psi}
\ee
for all sets $B\subseteq \RRR$ (with $1_B(\hat A)$ the projection to the subspace spanned by the eigenvectors of $\hat A$ with eigenvalue in $B$). Likewise, let $\mu^{\mc}_{\hat A}$ denote the probability distribution defined by the micro-canonical ensemble,
\be
\mu^{\mc}_{\hat A} (B) = \tr \bigl(1_B(\hat A) \, \hat\rho^{\mc}\bigr)
\ee
for all $B\subseteq \RRR$; $\mu^{\mc}_{\hat A}$ is the average of $\mu^\psi_{\hat A}$ with $\psi$ taken to be uniformly distributed in $\SSS(\Hilbert_{\mc})$.

Both MITE and MATE are ultimately of the following form (see also \cite{Sug07}): For a certain family $\sA$ of observables, consider the set of $\psi\in\SSS(\Hilbert_{\mc})$ 
for which $\mu^\psi_{\hat A}\approx \mu^{\mc}_{\hat A}$ for all $\hat A\in\sA$. MATE is obtained by taking $\sA=\sA_{\MATE}=\{\hat M_1,\ldots,\hat M_K\}$, and $\MITE_\ell$ by taking $\sA=\sA_{\MITE_\ell} = \cup_S \sA_S$ with the union taken over all regions in $\Lambda$ of diameter $\leq \ell$ and $\sA_S$ the set of \emph{all} self-adjoint operators on $\Hilbert_{S}$, more precisely,
\be\label{sASdef}
\sA_S = \bigl\{\hat A_0 \otimes \hat I_{S^c} : \hat A_0 \text{ self-adjoint on }\Hilbert_{S} \bigr\}\,,
\ee
where $\hat I$ denotes the identity operator and $S^c$ again the complement of $S$. Indeed, for $\hat A = \hat A_0 \otimes \hat I_{S^c}$ and $\psi\in\MITE_\ell$,
\be
\mu^{\mc}_{\hat A}(B)
= \tr\bigl( 1_B(\hat A) \, \hat\rho^{\mc}\bigr)
= \tr\bigl( 1_B(\hat A_0) \, \hat\rho_S^{\mc}\bigr)
\approx \tr\bigl( 1_B(\hat A_0) \, \hat\rho_S^\psi \bigr)
= \mu^\psi_{\hat A}(B)
\ee
for all $B\subseteq \RRR$ by \eqref{MITEdef}.

From this perspective it is obvious that MITE implies MATE when based on reasonable choices: Suppose that
\be\label{Lell}
\length \leq \ell_0\,,
\ee
where $\length$ is the length scale of the macro observables---the diameter of the cells $\Lambda_i$ on which the macro observables $\hat M_j$ were defined at the beginning of Section~\ref{sec:MATE}; that is, suppose that \eqref{rhopsirhomc} holds at least up to the length scale of the macro observables. This is commonly the case; e.g., for a cubic meter of gas at room conditions, we can realistically take $\length \approx 10^{-4}$ m and $\ell_0 \approx 10^{-3}$ m. 
As a consequence of \eqref{Lell}, $\sA_{\MATE} \subset \sA_{\MITE}$, so if $\psi\in\MITE$ then $\psi\in\MATE$.

{\it ETH.}---To come back to the eigenstate thermalization hypothesis (ETH), it comes in two variants: MATE-ETH and a more refined version MITE-ETH, according to whether the energy eigenstates are required to be in MATE or MITE. It is MITE-ETH that fails dramatically in some MBL systems, according to the findings of \cite{BAA06a,PH10,Imbrie}; there it is shown for certain MBL systems that a substantial fraction of the energy eigenstates (in a micro-canonical energy interval), or even all of them, lie outside of MITE. At the same time, it is easy to see that for every macroscopic quantum system (MBL or not), MATE-ETH must be almost satisfied, in the sense that most energy eigenstates $\ket{n} \in\SSS(\Hilbert_{\mc})$ are in MATE: Assuming that \eqref{dimeqdimmc} holds with $\varepsilon\ll \delta$, we obtain, writing $D=\dim\Hilbert_{\mc}$, that
\be\label{mosteigenstates}
\frac{1}{D} \sum_{n=1}^D \scp{n}{\hat P_{\eq}|n} = \frac{1}{D} \tr(\hat P_{\eq}) = 1-\varepsilon\,,
\ee
and since $\scp{n}{\hat P_{\eq}|n}$ cannot exceed 1, most of these terms must be close to 1.

If MATE-ETH holds strictly, i.e., if \emph{all} energy eigenstates in $\Hilbert_{\mc}$ are in MATE, then every state $\psi\in\SSS(\Hilbert_{\mc})$ will sooner or later reach MATE and spend most of the time in MATE in the long run. That is because \cite{GLMTZ09b}, writing $\overline{f(t)}=\lim_{T\to\infty} \frac{1}{T}\int_0^T f(t)\, dt$ for time averages, $\ket{n}$ for the energy eigenstate with eigenvalue $E_n$, and $\psi_t=e^{-i\hat Ht}\psi$,
\begin{align}
\overline{\scp{\psi_t}{\hat P_{\eq}|\psi_t}}
&= \sum_{n,n'} \scp{\psi}{n} \: \overline{e^{iE_nt} \scp{n}{\hat P_{\eq}|n'} e^{-iE_{n'}t}} \: \scp{n'}{\psi} \label{MATE-ETH-first}\\
&= \sum_{n} \bigl| \scp{\psi}{n} \bigr|^2  \scp{n}{\hat P_{\eq}|n} \\
&\geq \sum_{n} \bigl| \scp{\psi}{n} \bigr|^2  (1-\delta) \\
&=1-\delta\,,\label{MATE-ETH-last}
\end{align}
provided $\hat H$ is non-degenerate, i.e., $E_n\neq E_{n'}$ for $n\neq n'$ (using $\overline{e^{iEt}}=1$ if $E= 0$ and $=0$ otherwise).\footnote{In fact, the assumption of non-degeneracy can be dropped: If we number the eigenvalues as $E_n$ with $E_n\neq E_{n'}$ for $n\neq n'$ and let $\ket{n}$ denote the normalized projection of $\psi$ to the eigenspace of $E_n$, then the calculation \eqref{MATE-ETH-first}--\eqref{MATE-ETH-last} still applies.} Since its time average is close to 1, $\scp{\psi_t}{\hat P_{\eq}|\psi_t}$ must be close to 1 for most $t$ in the long run.

It follows from this that systems with MBL for which the transport coefficients vanish, so that an initial state $\psi$ with a non-uniform temperature will remain so indefinitely, cannot have all of its energy eigenfunctions in MATE. 
Since most energy eigenstates are in MATE, such $\psi$ must be a superposition of predominantly those rare eigenstates that are not in MATE.

This leads to the question whether there are macroscopic systems for which \emph{all} energy eigenstates are in MATE---i.e., whether MATE-ETH ever strictly holds.
It is known that this is so for a random Hamiltonian whose eigenbasis is uniformly chosen among all orthonormal bases \cite{GLMTZ09b}; see also \cite{vN29,GLTZ10}. Some numerical evidence \cite{KIH14} points to the existence of systems with realistic interactions for which all energy eigenstates are in MITE and thus also in MATE.

For MITE-ETH, there are several results \cite{RDO08,LPSW08,Rei10}, all of which assume
that the Hamiltonian is non-degenerate and has non-degenerate energy gaps, i.e.,
\be\label{noresonance}
E_{m}-E_{n} \neq E_{m'}-E_{n'} \text{ unless }
\begin{cases}\text{either } m= m' \text { and } n= n' \\
\text{or }m=n \text{ and }m'=n'\,,\end{cases}
\ee
a condition that is generically fulfilled. We note here two results, the first of which \cite{LPSW08} asserts that if all energy eigenstates in $\Hilbert_{\mc}$ are in MITE, then most $\psi\in\SSS(\Hilbert_{\mc})$ will sooner or later reach MITE and spend most of the time in MITE in the long run. More precisely, those $\psi$ will behave this way for which the effective number of significantly participating energy eigenstates is much larger than $\dim \Hilbert_S$ for any small $S$. The second result \cite{RDO08} shows that \emph{all} (rather than most) $\psi$ will ultimately reach MITE and stay there most of the time, under two assumptions, first again that all energy eigenstates are in MITE, and second Srednicki's \cite{Sre96,Sre99} extension of the ETH to off-diagonal elements, i.e., that for $\hat A\in\sA$ (here, $\sA = \cup_S \sA_S$ as in \eqref{sASdef}),
\be
\scp{m}{\hat A|n}\approx 0\text{ for }m\neq n
\ee
(see also \cite{Rei15}). Indeed, a calculation using \eqref{noresonance} shows that
\be\label{timevariance}
\overline{\biggl( \scp{\psi_t}{\hat A|\psi_t} - \overline{\scp{\psi_t}{\hat A|\psi_t}} \biggr)^2}
=\sum_{m\neq n} \bigl|\scp{\psi}{m}\bigr|^2 \; \bigl|\scp{m}{\hat A|n}\bigr|^2 \;
\bigl|\scp{n}{\psi}\bigr|^2\,,
\ee
and if $\bigl|\scp{m}{\hat A|n}\bigr|<\varepsilon\ll 1$ for all $m\neq n$, then the time variance \eqref{timevariance} is smaller than $\varepsilon^2$. If all $\ket{n}$ are in MITE, a calculation similar to \eqref{MATE-ETH-first}--\eqref{MATE-ETH-last} shows that
\be
\overline{\scp{\psi_t}{\hat A|\psi_t}} \approx \tr(\hat\rho^{\mc}\, \hat A)\,.
\ee
It follows that, for most $t$ in the long run, $\scp{\psi_t}{\hat A|\psi_t} \approx \tr(\hat\rho^{\mc}\, \hat A)$ for all $\hat A\in\cup_S\sA_S$ (in particular for projections), so $\psi_t\in \MITE$ for most $t$ in the long run.

{\it Mixed states.}---Once we have the notions of MITE and MATE for pure states $\psi$, they are easily generalized to mixed states $\hat \rho$: MATE occurs if $\tr(\hat P_{\eq}\, \hat\rho)>1-\delta$, and MITE if $\hat\rho_S \approx \hat\rho^{\mc}_S$ for all subsystems $S$ defined by spatial regions of diameter $\leq \ell_0$. Note that neither MATE nor MITE requires that $\hat\rho$ be close to $\hat\rho^{\mc}$ or $\hat\rho^{(\beta)}$.

{\it Thermal equilibrium in classical mechanics.}---Only 
one of the two notions MITE and MATE
can be satisfied for pure states in classical mechanics, namely MATE. That is because a ``pure state'' corresponds in classical mechanics to a point $X$ in phase space, while a ``mixed state'' corresponds to a probability distribution over phase space. Since $X$ specifies the positions and momenta of all particles, it also provides a pure state for any subsystem.
In contrast, in quantum mechanics $\hat{\rho}^\psi_S$ can be mixed, and in fact is mixed except for product states. So in classical mechanics it is never true for a system in a pure state that a subsystem $S$ could have a state close to a thermodynamic ensemble such as the marginal (obtained by integrating out the variables not belonging to $S$) of the micro-canonical distribution (i.e., uniform over the energy shell) or the canonical one for the whole system. 
In contrast, MATE is analogous to Boltzmann's \cite{B96,Gol99,GL,L07} notion of thermal equilibrium for a closed classical system, based on a partition of phase space into macro states $\Gamma_\nu$. (Note that there is no difference between MATE and TMATE classically, as all observables commute.)

{\it Abstract MITE.}---A natural mathematical generalization that is often interesting to consider is based on dropping the idea that $S$ corresponds to a region in 3-space and regarding $S$ as an abstract subsystem defined by any splitting of Hilbert space into a tensor product,
\be
\Hilbert_{\mc} \subseteq \Hilbert_S \otimes \Hilbert_{S^c}\,,
\ee
where $S^c$ can be thought of as just an index for another Hilbert space. For example, $S$ may comprise the spin degrees of freedom and $S^c$ the position degrees of freedom, or $S$ may comprise the oxygen atoms and $S^c$ all other atoms in the system. Given a list of subsystems $S_1,\ldots,S_r$, one can define
\be\label{MITElistdef}
\MITE_{S_1,\ldots,S_r} = \bigcap_{i=1}^r\Bigl\{ \psi\in\SSS(\Hilbert_{\mc}): \hat\rho_{S_i}^\psi \approx \hat\rho^{\mc}_{S_i} \Bigr\}\,.
\ee
Canonical typicality implies that, if $r$ is not too large and each $S_i$ is not too large (a sufficient condition is $\dim\Hilbert_{S_i} \ll \sqrt{\dim \Hilbert_{\mc}}$), then
most $\psi\in\SSS(\Hilbert_{\mc})$ are in $\MITE_{S_1,\ldots,S_r}$; see Theorem~1 in \cite{PSW05,PSW06} and Proposition~1 in \cite{GHLT15b} for a precise and quantitative statement of this fact.

One can also consider the set $\MITE_{\text{most}}$ comprising those $\psi\in\SSS(\Hilbert_{\mc})$ for which $\hat{\rho}_S^\psi \approx \hat\rho^{\mc}_S$ holds for \emph{most} abstract subsystems $S$ of dimension $\leq d_0$. If $d_0\ll \sqrt{\dim\Hilbert_{\mc}}$, then also $\MITE_{\text{most}}$ has most of the surface area of $\SSS(\Hilbert_{\mc})$ \cite{GHLT15b}.
On the other hand, given any pure state $\psi\in\SSS(\Hilbert_{\mc})$, $\hat{\rho}_S^\psi \approx \hat\rho^{\mc}_S$ cannot hold for \emph{all} abstract subsystems $S$ of dimension $\leq d_0$ simultaneously \cite{Lych,GHLT15b}.

\section{Conclusions}

Perhaps the most surprising aspect of the situation is that the various criteria for thermal equilibrium of pure states proposed in the literature 
fall into two groups that differ substantially in how much they demand.

Arguably, the essence of thermal equilibrium is what characterizes it in thermodynamics: that a system appears stationary on the macro level, and that temperature and all chemical potentials are spatially uniform. This corresponds to MATE, which may therefore be regarded as the direct expression of thermal equilibrium.
On the other hand, since MITE is the stronger statement, and since it is usually true that macroscopic quantum systems approach MITE (MBL systems being an exception), it is natural to consider MITE, and it would seem artificial to not regard it as a new kind of thermal equilibrium property emerging from quantum entanglement.

\bigskip

\noindent\textit{Acknowledgments.}
J.~L.~Lebowitz was supported in part by the National Science Foundation [grant  DMR1104500].

\end{document}